\pdfoutput=1
\documentclass[twoside]{amsart}
\usepackage[norelsize]{algorithm2e}

\usepackage{dirtytalk}
\usepackage{url}
\usepackage[a4paper, portrait, margin=0.5in]{geometry}
\usepackage{multicol}
\usepackage{tipa}
\usepackage{setspace}
\usepackage{caption}
\usepackage{amsmath,amsthm,amssymb}

\title{Automatic Wikipedia Link Generation Based On Interlanguage Links}
\author{Michael Lotkowski \\ \href{mailto:s1207068@sms.ed.ac.uk}{s1207068@sms.ed.ac.uk}}

\begin{document}
\onehalfspacing
\maketitle
\nocite{*}
\begin{multicols}{2}
\begin{abstract}
   This paper presents a new way to increase interconnectivity in small Wikipedias (fewer than a $100,000$ articles), by automatically linking articles based on interlanguage links. Many small Wikipedias have many articles with very few links, this is mainly due to the short article length. This makes it difficult to navigate between the articles. In many cases the article does exist for a small Wikipedia, however the article is just missing a link. Due to the fact that Wikipedias are translated in to many languages, it allows us to generate new links for small Wikipedias using the links from a large Wikipedia (more than a $100,000$ articles).
\end{abstract}
\section{Introduction}
Wikipedia is a free-access, free-content Internet encyclopaedia, supported and hosted by the non-profit Wikimedia Foundation. Those who can access the site can edit most of its articles. Wikipedia is ranked among the ten most popular websites and constitutes the Internet's largest and most popular general reference work \cite{wikipedia}. Like most websites, it's dominated by hyperlinks, which link pages together. Small Wikipedias (fewer than a $100,000$ articles) have many articles with very few links, this is mainly due to the short article length. This makes it difficult to navigate between the articles. In many cases the article does exist for the small Wikipedia, however the article is just missing a link. This will also be true for other small Wikipedias. Wikipedia provides an option to view the same article in many languages. This is a useful option if the user knows more than one language, however most users would prefer to read an article in the same language. This leads me to believe that using a large Wikipedia (more than a $100,000$ articles), it is possible to automatically generate links between the articles.
	The problem of automatically generating links is a difficult problem as it requires a lot of semantic analysis to be able to match articles that are related.

\section{Methodology}
For the automatic page linking two datasets are required; a small Wikipedia and a large Wikipedia, which we will use to supplement the links for the small Wikipedia.
The large size of the English Wikipedia ($\sim5,021,159$ articles)\cite{wikipedia_stats} makes it an ideal set to draw links from. The Scots language Wikipedia has $\sim35,300$ articles\cite{wikipedia_stats}, and it's not very well connected, with the strongly connected component covering only $56.72 \%$ of all nodes, thus it makes it an ideal candidate to be used as the small Wikipedia. 
Wikimedia provides database dumps for every Wikipedia \cite{wikimedia_dumps}. First, we need to extract all links from the dumps using Graphipedia \cite{graphipedia}
. This will generate a neo4j database, containing all pages as nodes and links as edges. I have also altered Graphipedia to output a json file to make it smaller and easier to manipulate in Python \cite{graphipedia-michael}. After the preprocessing is done, then we will run analysis on the base graph of the small Wikipedia. Afterwards we will enrich the base graph of the small Wikipedia using the graph structure of the large Wikipedia, and run the same analysis. Lastly, we will combine the two graphs together and run the same analysis. The enriched graph will have to be normalised as it's not possible to get new links for all pages.

To evaluate this method, I will compare the size of the strongly connected component, the average number of links and the number of pages which have no outgoing links. For this method to be successful it has to have a significant increase in interconnectivity and significantly decrease dead-end pages. The viability of the method will also depend on it's running time performance.

The strongly connected component is important here as we want the user to be able to navigate between any pair of articles on the network. The larger the component the more articles are accessible, without the need of manual navigation to an article. I will determine the strongly connected component by using an algorithm presented in \cite{Tarjan72depthfirst}.

\section{Existing Research}
There are many automatic link generation strategies for Wikipedia articles \cite{Granitzer2008e}, however they are all trying to generate links based on the article's content. This is a good approach for the large Wikipedias wherein the articles are comprehensive. This approach underperforms on small Wikipedias where we see a lot of short articles. In the small Wikipedia, in many cases the articles that are related do exist, however they are not linked. Many strategies involve machine learning \cite{vsolc2008automatic} \cite{ruiz2011accessing}, which can be a complex and time expensive task.
Through my research I was not able to find any similar methods to the one presented.

\section{Solution}
Wikimedia provides an SQL dump of interlanguage mappings between articles. Combined with the page titles dump, we can join the two to create a mapping between the Scots and English titles. For each Scots title we look up the equivalent English title. Then search the English Wikipedia for its the first degree neighbours. Subsequently for each neighbour node, we look up the Scots title equivalent. If it exists, then create an edge between the initial Scots title and the Scots title equivalent of the English neighbour node. Below is a pseudo code outlining this algorithm.
 \hfill \newline

\begin{algorithm}[H]
 \KwData{SW: small Wikipedia Dump}
 \KwData{LW: large Wikipedia Dump}
 \KwData{M: Mapping between page titles}
  \KwData{G: New graph}
 \KwResult{A small Wikipedia graph with links mapped from the large Wikipedia}
	
 \For{page in SW}{
 	large\_page = M.toLW(page)
  	links = LW.linksForPage(large\_page)\\
  	\For{link in links}{
  		\If{link in M}{
  			G[page].addLink(M.toSW(link))
  		}
  	} 
  }
  
  \Return G
  \hfill \newline
 \caption{Automatically generate new links}
\end{algorithm}
\hfill \newline
This algorithm has worst case complexity of $O(nl)$ where $n$ is the number of pages in the small Wikipedia, and $l$ is the number of links in the large Wikipedia. However on average it performs in $\Omega(n)$, as the number of links per page is relatively small, on average about 25 per page in the English Wikipedia \cite{sixdegrees}. Formally we can define it as:

  \begin{enumerate}
\item Let the graph $G_S = (V_S, E_S)$ be the graph of the small Wikipedia
\item Let the graph $G_L = (V_L, E_L)$ be the graph of the large Wikipedia
\item Let the graph $G = (V, E)$ be the newly constructed graph
\item Let the function $SL(x)$ be a mapper such that $SL(x \in V_S) \to x' \in V_L$
\item Let the function $SL(x)$ be a mapper such that $LS(x \in V_L) \to x' \in V_S$
\item Let $V = V_S$

\item For $v \in V_S$:
 
{\setlength\itemindent{25pt} \item For $(SL(v), u) \in E_L$:
 {\setlength\itemindent{50pt}\item $E = u \cup E_S$}}

\end{enumerate}

After we have built this graph we can combine it, simply by adding the edges from the newly formed graph to the vanilla graph, where those edges do not exist.

\section{Results}

The vanilla Scots refers to the unaltered graph of the Scots Wikipedia, the enriched Scots is the generated graph of the Scots Wikipedia based on the English Wikipedia and the combined Scots is the combination of the two previous graphs.

The enriched Scots graph has only 31516 nodes, as not all pages have a direct English mapping. To normalise the graph I have added the missing 13154 nodes without any links, so that the below results are all out of 44670 nodes of the total graph.

The main objective of the algorithm was to maximise the size of the SCC. Table \ref{table:scc} presents the results.
\hfill \newline

\begin{center}
\captionof{table}{Size of the SCC (44670 nodes total)}
\begin{tabular}{l|cc}
  Graph type & Nodes in SCC & \%\\
  \hline
  Vanilla Scots & 25338 & 56.72\\
  Enriched Scots & 27366 & 61.26\\
  Combined Scots & 36309 & 81.28

\end{tabular}
\label{table:scc}
\end{center}

\hfill \newline

Even the enriched Scots graph has a larger SCC than the vanilla graph, making it better connected, and thus easier to navigate. The combined graph sees an improvement of $43.30 \%$ in the size of SCC compared to the vanilla graph.

\hfill \newline \\

\begin{center}
\captionof{table}{Average degree of a node}
\begin{tabular}{l|c}
  Graph type & Avg. Degree\\
  \hline
  Vanilla Scots & 14.91\\
  Enriched Scots & 26.83\\
  Combined Scots & 34.84 

\end{tabular}
\label{table:degree}
\end{center}

\hfill \newline

Table \ref{table:degree} shows the average amount of links that an article has. The combined Scots graph has increased the average degree by $133.67 \%$ compared to the vanilla Scots graph, which means the users have on average more than twice the amount of links to explore per page.

\hfill \newline

\begin{center}
\captionof{table}{Deadend articles (44670 nodes total)}
\begin{tabular}{l|cc}
  Graph type & Nodes in SCC & \%\\
  \hline
  Vanilla Scots & 456 & 1.02\\
  Enriched Scots & 13537 & 30.30\\
  Combined Scots & 218 & 0.49

\end{tabular}
\label{table:deadends}
\end{center}

\hfill \newline

Table \ref{table:deadends} shows the number of articles that have no outgoing links. The enriched Scots graph had to be normalised by adding 13154 nodes with degree 0. The combined Scots graph sees an improvement of $52.19 \%$ compared to the vanilla Scots graph. This result is significant as now only less than half a percent of pages will result in a repeated search or manual navigation.

\section{Conclusion}

From the results presented in the previous section, we see a significant improvement in the connectivity of the article network. The method presented here is deterministic and of low complexity, thus making it a viable option to enrich even large Wikipedias which have millions of nodes and many millions of links. This method can be applied even to languages that use a non latin character set as Wikipedia provides a mapping between any pair of languages. For this reason, it is also possible to apply this algorithm in an iterative manner, sourcing links from many languages.

A drawback of this approach, however, is that we can't integrate the generated links within the article text, thus the links would have to be displayed separately on the site. The effectiveness of this approach also depends on the size ratio of the used Wikipedias. 

Since this is an additive method, thus the combined graph will be at least as well connected as the vanilla graph.

Because of the increase in interconnectivity between the articles, the small Wikipedia's users will be able to explore a more significant portion of their Wikipedia and also incline them to expand the articles to accommodate the extra links. This method also benefits machine learning tasks, as it would be easier to create artificial agents which consume the knowledge from a small Wikipedia. Whereas currently the small Wikipedias are poorly linked making it difficult to run semantic analysis. 

\section{Further Improvements}
Given more time I would want to run it against every Wikipedia and use pagerank to determine the best links for the article. This is not feasible as running pagerank on the English wikipedia will take a very long time. 
I would also like to explore the possibility of retrieving snippets from a large Wikipedia and then suggesting them or automatically translating them for the small Wikipedia to allow automatic content creation.

\end{multicols}
\end{document}